\documentclass[a4paper]{article}
\usepackage{tikz}
\usepackage{pgfplots}
\usepackage[skip=5pt]{caption}

\newcommand{\figcap}{\setlength{\baselineskip}{.5em}}

\usepackage{INTERSPEECH2018}

\title{Do WaveNets Dream of Acoustic Waves?}
\name{Kanru Hua}
\address{University of Illinois, U.S.A.}
\email{khua5@illinois.edu}

\begin{document}

\maketitle

\begin{abstract}
  Various sources have reported the WaveNet deep learning architecture being able to generate high-quality speech, but to our knowledge there haven't been studies on the interpretation or visualization of trained WaveNets. This study investigates the possibility that WaveNet understands speech by unsupervisedly learning an acoustically meaningful latent representation of the speech signals in its receptive field; we also attempt to interpret the mechanism by which the feature extraction is performed. Suggested by singular value decomposition and linear regression analysis on the activations and known acoustic features (e.g. F0), the key findings are (1) activations in the higher layers are highly correlated with spectral features; (2) WaveNet explicitly performs pitch extraction despite being trained to directly predict the next audio sample and (3) for the said feature analysis to take place, the latent signal representation is converted back and forth between baseband and wideband components.
\end{abstract}

\section{Introduction}

Until fairly recently, researches on the digital synthesis of speech signals have mainly been focusing on the manipulation of manually-engineered audio and speech features. With the advent of deep neural network (DNN) architectures directly modeling the waveform \cite{vandenoord-2016, tokuda-2015}, it is now possible to construct systems automating a significant portion of the signal processing tasks in speech analysis-synthesis.

In particular, the WaveNet architecture \cite{vandenoord-2016} and several recently proposed WaveNet-derived convolutional networks \cite{arik-2017, shen-2017} have demonstrated near-natural quality ratings in single-speaker speech synthesis. However, the high computational demand for training and evaluating the said networks remains a limiting factor on the applications. In addition, attempts on certain tasks such as multi-speaker speech synthesis \cite{arik-2017} and denoising \cite{rethage-2017} still show opportunities for enhancements. It is desirable to gain an understanding of the learned models to set forth possible topological changes and other improvements on the networks as well as training techniques.

It is well-known that the weights in first layer of a trained convolution neural network often reveal interpretable features \cite{lecun-1989}. For the visualization of features in the subsequent layers, various approaches have been proposed to find a representative input that maximizes the activation of a particular hidden unit by means of gradient descent or deconvolutional networks \cite{zeiler-2014, samek-2017}. A similar class of methods that relates the learned network to features in the input space is to assign a local measure of how much the variation of the input affects the final output, along the input dimensions for a given output \cite{montavon-2017}. Despite the reported successes of pattern discovery in the weight or input space on image recognition/classification models, it is not clear on how to meaningfully apply such methods to time-series prediction, especially when a visualization in the frequency domain is desired for its relevance to human perception.

Another related area is the interpretation of recurrent neural networks (RNN), since the dilated module outputs in a WaveNet are also time series, similar to the hidden states in a RNN. Empirical studies in the said area have mostly focused on the task of language modeling: Strobelt, et al.~\cite{strobelt-2018} designed a tool for inspecting reoccuring hidden state patterns associated with a subsequence from the inputs; Karpathy, et al.~\cite{karpathy-2015} tested the hypothesis that the hidden cells contain interpretable information by color-coding the input text according to a sequence of activation values. However, our initial attempt to apply such pattern discovery methods on WaveNet was hindered by the large amount of data generated by the network due to the sampling rate and number of layers much greater than a typical RNN setup for natural language processing tasks.

This study attempts to answer the following questions: whether or not a WaveNet model trained on speech data \textit{understands} the signal by performing feature extraction, and how is feature extraction done inside the trained network. To circumvent the difficulties listed above, we hypothesize that WaveNet is able to \textit{unsupervisedly} learn features (e.g. fundamental frequency and band energy) amenable to human understanding, and to verify this hypothesis we train linear mappings from the activations of a layer to features extracted from the input signals. This paper starts with a brief review of the original WaveNet architecture in section 2; the experiment setup is described in section 3 and the results are discussed in section 4. We conclude the study by summarizing the findings and implications for future works in section 5.

\section{WaveNet}

WaveNet \cite{vandenoord-2016} is an autoregressive discrete-time signal modeling network that predicts the next sample by conditioning on a certain number of previously generated samples. To synthesize speech from text inputs, the prediction is conditioned on an extra set of variables containing the linguistic context.
\begin{figure}[htb]
\centerline{\includegraphics[width=6cm]{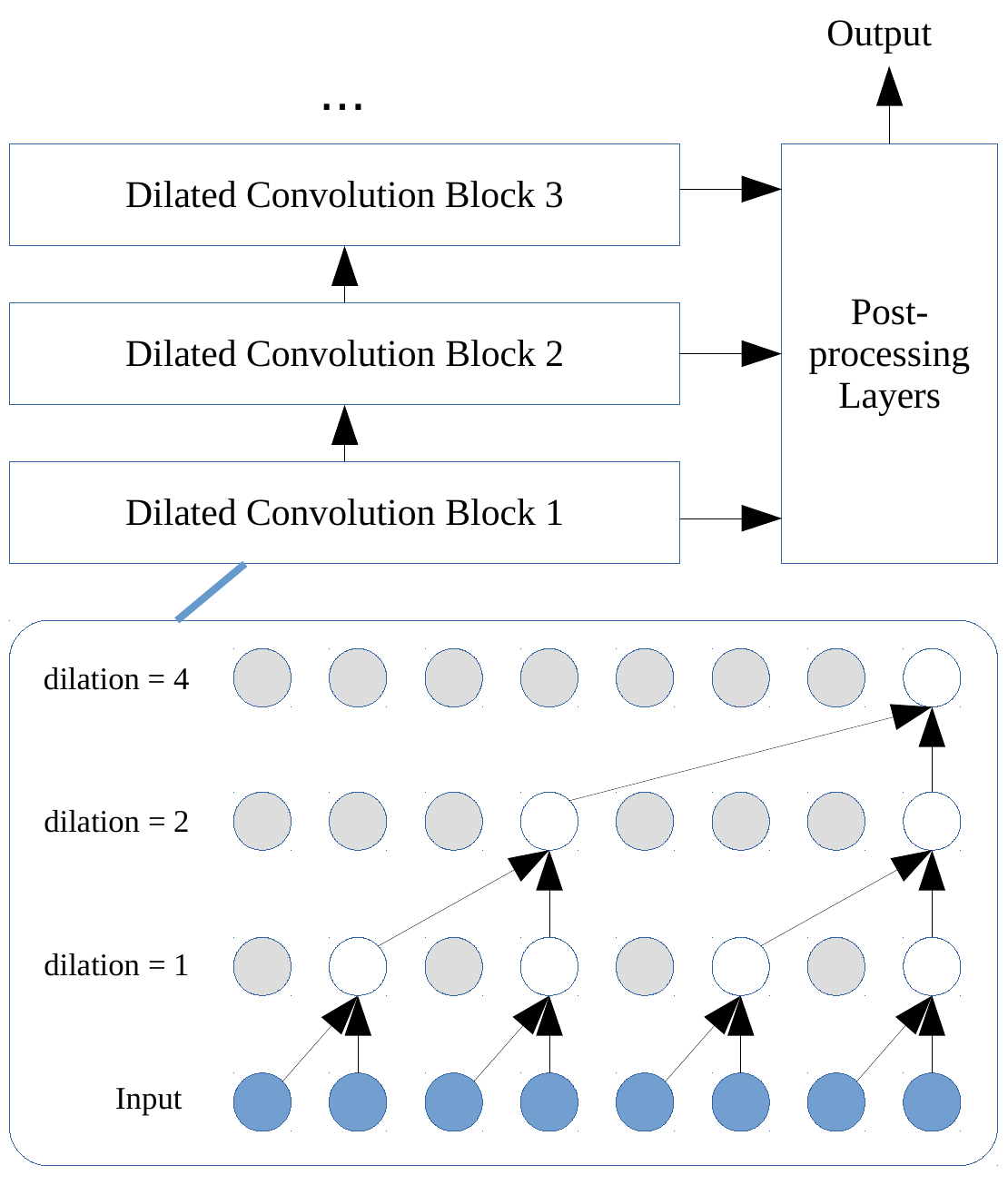}}
\caption{\figcap{A simplified overview of the WaveNet architecture. The bottom panel outlines the data flow in the first dilated convolution block in each time step.}}
\label{fig:wavenet-arch}
\vspace{-10pt}
\end{figure}
Figure~\ref{fig:wavenet-arch} shows the model structure implementing the said computation. The main element for computation is a set of dilated convolution layers, each of which defines the following operation,
\begin{align}
  \mathbf y[t] = \sum_{n=0}^N W_n \mathbf x[t - nk]
\end{align}
where $\mathbf x$ and $\mathbf y$ are the input and the output sequences of vectors; $W$ is the tensor for weights; $k$ is the dilation factor and $N$ is the convolution size. In the original WaveNet setup, $N$ is set to 1 for every layer and $k$ grows as an integer power of 2 so that the topology becomes effectively a binary tree. The same block consisting of dilated convolution layers of exponentially growing sizes is repeatedly cascaded for a few times, reaching a combined receptive field covering a few hundred milliseconds.

Each layer also implements an activation gating operation combining the results from two dilated convolutions and a residual connection. The input to the next layer can be computed as,
\begin{align}
  \mathbf y = \mathbf x + \mathrm{tanh}\left( W_f * \mathbf x + V_f \mathbf c \right) \sigma\left( W_g * \mathbf x + V_g \mathbf c \right)
  \label{eq:wavenet-layer}
\end{align}
where $W_f$ and $W_g$ are the filtering and gating tensors. Optionally, the tensors $V_f$ and $V_g$ can incorporate an additional input sequence $\mathbf c$ that carries the linguistic information.

To estimate a conditional distribution on the waveform amplitude, the speech data is quantized to 8 bits and the output layer is formulated as a softmax function over 256 categories. The outputs from the convolution layers are combined by a post-processing block consisting of a few fully-connected layers. Quantization artifacts can be reduced by a non-linear quantization scheme \cite{itut-1988}.

\section{Experiment Setup}

We hypothesize that at each time step, interpretable speech features can be derived from linear combinations of the outputs of a dilated convolution layer, since matrix multiplication is one of the elementary operations in a convolutional neural network. The relation between the latent representation formed in a WaveNet and speech features can be tested by regression analysis with a properly chosen objective function.

\subsection{WaveNet Models}

To eliminate correlations caused by the features' dependency on linguistic inputs, this study specifically investigates unconditioned WaveNet models, i.e., the case when $\mathbf c = 0$ in (\ref{eq:wavenet-layer}). Speaker-dependent models with the identical configuration (Table~\ref{tab:model-config}) are trained on four speakers (2 male and 2 female) from the CMU Arctic speech database \cite{kominek-2004}.
\begin{table}[h]
\centering
\caption{\figcap{Model and Training Configuration}}
\begin{tabular}{|l|l|}
  \hline
    Sample rate & $16$ kHz \\ \hline
    Layer size & $64$ units \\ \hline
    Dilation in each block & $1, 2, 4, 8, 16, 32, 64, 128, 256, 512$ \\ \hline
    Number of blocks & $5$ (total receptive field of $320$ ms) \\ \hline
    Optimization & Adam, learning rate $= 0.001$ \\ \hline
    Training set & $1052$ sentences, around $1$ hour \\ \hline
    Test set & $80$ sentences \\ \hline
\end{tabular}
\label{tab:model-config}
\vspace{-10pt}
\end{table}

For each speaker, the trained WaveNet is evaluated on the test set and the activations from all hidden layers are stored for regression analysis; regression models are trained on 60 sentences from the test set and evaluated on the rest 20 sentences. Unless otherwise noted, the regression linearly projects a 64-dimensional activation vector (according to the layer size) with an extra bias term to the target feature space.

\subsection{Features}

This study considers the following types of speech features: speech waveform (prior to $8$-bit quantization), fundamental frequency (F0), band energy (dB), and STFT magnitude spectrogram (both wideband and narrowband).


The list includes both rapidly varying features (waveform and wideband spectra) and quasi-stationary features (F0, band energy and narrowband spectra). Details on computing the target features are listed as the follows.

\subsubsection{Details on Feature Extraction}

For the reconstruction of waveform, it is important to consider both input and target signals, which differ by an 1-sample offset and would respectively indicate the representation's capability at linearizing the discrete data and predicting the next sample.

Log fundamental frequency is extracted from the input data using adYANGsaf \cite{hua-2017}. Our preliminary experiments showed that warping the frequency in linear or log scale does not significantly affect the results.

Without assuming that WaveNet learns an energy representation on an auditory frequency scale, we extract band energy from the input signal using a 2nd-order Butterworth filterbank with 20 channels uniformly spanned from $0$ to $8$ kHz. Preliminary experiments showed that a higher Spearman's correlation coefficient is achieved using a log-scale energy representation.

Finally, we attempt to reconstruct STFT magnitude spectra from activation vectors at the output end of each layer. Wideband spectra are taken using a $4$ ms Blackman window and the window length for narrowband spectra is $32$ ms.

\subsubsection{Objectives for Regression}

Ordinary least squares (OLS) regression is applied to reconstruct waveform, log F0, and log band energy features. Following a recent study on directly modeling magnitude spectrum using DNNs \cite{takaki-2017}, we formulate the problem of magnitude spectra estimation as minimizing the KL divergence. However in the case of this study, Itakura-Saito distance based regression was found to better discriminate voiced and unvoiced speech. The said regression can be easily done in 2 Newton iterations, initialized by an OLS regression on log magnitude spectra.

\section{Findings and Discussions}

A theoretical limit on the accuracy of feature reconstruction can be derived by comparing the receptive field at a given layer to the minimal duration required to characterize a certain feature. The result is a useful reference in the following analysis when characterizing the change in accuracy across layers. For example, a reliable estimation of the fundamental frequency requires at least two periods of time-domain signal, which correspond to around 160 samples for female voices at a $16$ kHz sample rate. So it is expected that the characterization of F0 will appear no earlier than layer 7 ($\log_2 160 \approx 7.32$). Similar deductions can be made on other features.

Due to limited space, some figures shown in this section only include speaker \textit{slt} (female) and \textit{bdl} (male). The author has verified that similar trends can be observed on the rest of the speakers being tested.

\subsection{Waveform Regression from Hidden Layers}

\begin{figure}
\centering
\input{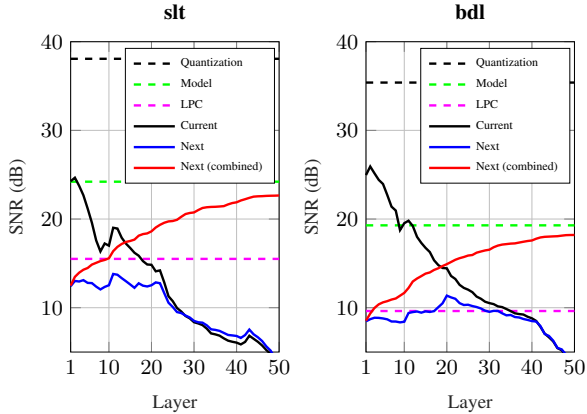}
\caption{\figcap{Signal-to-noise ratios of waveforms linear-transformed from WaveNet activations. Dashed lines represent reference levels; solid lines represent the regression results. Although the layer-wise SNR decreases, the multi-layer prediction becomes more accurate as more layers are involved.}}
\label{fig:waveform}
\vspace{-10pt}
\end{figure}

Figure~\ref{fig:waveform} shows the signal-to-noise ratio, measured in decibels, of both current-sample and next-sample reconstructions from the activations. For predicting the next sample, we also tested larger regression models with all the activations below a certain layer as the input. The SNR due to quantization effects, the final output from the model, and a 512-order linear predictive filter representing the best linear autoregressive model are plotted for reference.

It is seen that, with the exception of the first few layers, the single-layer waveform prediction/reconstruction achieves a SNR only comparable to high-order linear prediction and the performance decreases in upper layers. However, within a mere $2$ dB margin, the combined prediction based on activations from all layers approaches the model's final prediction. Such a trend suggests the possibility that the dilated convolution layers transform the raw signal into a more abstract representation and complementary information is derived at different layers. Further evidences can be gathered in the following analysis on slowly varying features.

\subsection{F0 Regression from Hidden Layers}
\label{sec:logf0}

In agreement with the theoretical limit derived in the beginning of this section, a significant increase in Spearman's correlation coefficient on log F0 can be observed for all speakers near layer 7 shown in Figure~\ref{fig:f0}. Notably, the first plateau is reached one layer later in the case of speaker \textit{jmk}, who has the lowest average F0. The consistent trend that a high correlation can be reached at a layer index close to the theoretical limit implies the importance of pitch estimation to feature analysis within the WaveNet model.

The correlation coefficient peaks around 0.9 after reaching a second plateau starting between layer 15 and 17, which belong to the second convolution block (see Figure~\ref{fig:wavenet-arch} for a graphical illustration). Considering that WaveNet maintains a reasonably good reconstruction of the input waveform in the second convolution block (Figure~\ref{fig:waveform}), it can be inferred that a refined F0 analysis is performed in the second half of the block. This claim is also supported by the fact that the dilation factors in layers between 15 and 17 correspond to the $125$ - $500$ Hz frequency range, which relates to the first one or two speech harmonics.

\begin{figure}
\centering
%
%
\definecolor{mycolor1}{rgb}{0.00000,0.44700,0.74100}%
\definecolor{mycolor2}{rgb}{0.85000,0.32500,0.09800}%
\definecolor{mycolor3}{rgb}{0.92900,0.69400,0.12500}%
\definecolor{mycolor4}{rgb}{0.49400,0.18400,0.55600}%
\begin{tikzpicture}

\begin{axis}[%
width=2.3in,
height=1.3in,
at={(0.517in,0.2in)},
scale only axis,
xmin=1,
xmax=50,
xlabel near ticks,
xtick = {1, 10, 20, 30, 40, 50},
xlabel style={font=\color{white!15!black}},
xlabel={\footnotesize{Layer}},
ymin=0.2,
ymax=1,
ylabel near ticks,
ylabel style={font=\color{white!15!black}},
ylabel={\footnotesize{Spearman's Correlation Coef.}},
axis background/.style={fill=white},
xmajorgrids,
ymajorgrids,
legend style={at={(0.97,0.03)}, anchor=south east, legend cell align=left, align=left, legend plot pos=left, draw=black}
]
\addplot [color=mycolor1, line width=1.0pt]
  table[row sep=crcr]{%
1	0.227175340056419\\
2	0.233183413743973\\
3	0.250015765428543\\
4	0.295739889144898\\
5	0.37185674905777\\
6	0.448362201452255\\
7	0.624115347862244\\
8	0.698698699474335\\
9	0.76671290397644\\
10	0.772530257701874\\
11	0.767242312431335\\
12	0.772384941577911\\
13	0.77683436870575\\
14	0.791667759418488\\
15	0.839769065380096\\
16	0.862636804580688\\
17	0.875671684741974\\
18	0.882828652858734\\
19	0.885236620903015\\
20	0.885654628276825\\
21	0.887519776821136\\
22	0.883828282356262\\
23	0.880689918994904\\
24	0.890211284160614\\
25	0.904799997806549\\
26	0.914158046245575\\
27	0.905923664569855\\
28	0.89944189786911\\
29	0.891347706317902\\
30	0.882738173007965\\
31	0.875872075557709\\
32	0.872485816478729\\
33	0.866908490657806\\
34	0.85919064283371\\
35	0.852692008018494\\
36	0.848706245422363\\
37	0.845547795295715\\
38	0.841897547245026\\
39	0.836613893508911\\
40	0.833506166934967\\
41	0.832359313964844\\
42	0.832690358161926\\
43	0.8293097615242\\
44	0.822085022926331\\
45	0.820445597171783\\
46	0.814378917217255\\
47	0.802896678447723\\
48	0.783024907112122\\
49	0.759679138660431\\
50	0.757195055484772\\
};
\addlegendentry{slt}

\addplot [color=mycolor2, line width=1.0pt]
  table[row sep=crcr]{%
1	0.30616956949234\\
2	0.313838213682175\\
3	0.340088307857513\\
4	0.389275729656219\\
5	0.483766406774521\\
6	0.573037147521973\\
7	0.629016399383545\\
8	0.747399508953094\\
9	0.757488131523132\\
10	0.765439987182617\\
11	0.764316618442535\\
12	0.754601418972015\\
13	0.758980214595795\\
14	0.766949057579041\\
15	0.761112034320831\\
16	0.769745528697968\\
17	0.793720662593842\\
18	0.83470231294632\\
19	0.846903145313263\\
20	0.845368623733521\\
21	0.843870639801025\\
22	0.840985238552094\\
23	0.838360786437988\\
24	0.83842670917511\\
25	0.843596577644348\\
26	0.853296279907227\\
27	0.857245802879333\\
28	0.86655992269516\\
29	0.863822162151337\\
30	0.860028743743896\\
31	0.857441186904907\\
32	0.849776387214661\\
33	0.849663019180298\\
34	0.848271310329437\\
35	0.847886860370636\\
36	0.866545557975769\\
37	0.882681548595428\\
38	0.879498898983002\\
39	0.887826502323151\\
40	0.889315068721771\\
41	0.888538420200348\\
42	0.890730500221252\\
43	0.893465638160706\\
44	0.898679494857788\\
45	0.894173562526703\\
46	0.901364147663116\\
47	0.911548852920532\\
48	0.910425305366516\\
49	0.904616296291351\\
50	0.901093780994415\\
};
\addlegendentry{bdl}

\addplot [color=mycolor3, line width=1.0pt]
  table[row sep=crcr]{%
1	0.313013941049576\\
2	0.365515172481537\\
3	0.395274251699448\\
4	0.423867255449295\\
5	0.469160050153732\\
6	0.5229572057724\\
7	0.559837639331818\\
8	0.634588956832886\\
9	0.683734834194183\\
10	0.696907818317413\\
11	0.693902611732483\\
12	0.690712153911591\\
13	0.690979480743408\\
14	0.686717867851257\\
15	0.689681529998779\\
16	0.703800201416016\\
17	0.736329913139343\\
18	0.834602415561676\\
19	0.879916131496429\\
20	0.873478829860687\\
21	0.868002891540527\\
22	0.864700376987457\\
23	0.854373276233673\\
24	0.85400253534317\\
25	0.86586856842041\\
26	0.883680820465088\\
27	0.893957376480103\\
28	0.896586954593658\\
29	0.899884402751923\\
30	0.898484289646149\\
31	0.895874917507172\\
32	0.892736673355103\\
33	0.892553508281708\\
34	0.896354794502258\\
35	0.901847004890442\\
36	0.909155070781708\\
37	0.91294914484024\\
38	0.902728796005249\\
39	0.891767263412476\\
40	0.884282767772675\\
41	0.876001179218292\\
42	0.862781465053558\\
43	0.849720299243927\\
44	0.831838369369507\\
45	0.821137309074402\\
46	0.814477443695068\\
47	0.807862043380737\\
48	0.795222699642181\\
49	0.787975192070007\\
50	0.790252983570099\\
};
\addlegendentry{jmk}

\addplot [color=mycolor4, line width=1.0pt]
  table[row sep=crcr]{%
1	0.294399410486221\\
2	0.323537975549698\\
3	0.356461942195892\\
4	0.403104871511459\\
5	0.487942516803742\\
6	0.565259337425232\\
7	0.659889578819275\\
8	0.726841449737549\\
9	0.770501017570496\\
10	0.78350430727005\\
11	0.777578234672546\\
12	0.779462695121765\\
13	0.783189296722412\\
14	0.799280047416687\\
15	0.817892670631409\\
16	0.84694504737854\\
17	0.929633438587189\\
18	0.946480214595795\\
19	0.955144464969635\\
20	0.955900967121124\\
21	0.955204248428345\\
22	0.954307317733765\\
23	0.951181411743164\\
24	0.952991306781769\\
25	0.956055223941803\\
26	0.959305703639984\\
27	0.947323739528656\\
28	0.947959899902344\\
29	0.940812349319458\\
30	0.935734868049622\\
31	0.931904196739197\\
32	0.92897629737854\\
33	0.926640272140503\\
34	0.92374062538147\\
35	0.92095011472702\\
36	0.919446527957916\\
37	0.917267978191376\\
38	0.91562032699585\\
39	0.914662897586823\\
40	0.913738667964935\\
41	0.91330486536026\\
42	0.91222620010376\\
43	0.910759925842285\\
44	0.907972514629364\\
45	0.907216846942902\\
46	0.890942752361298\\
47	0.880541741847992\\
48	0.870132684707642\\
49	0.843166649341583\\
50	0.836246371269226\\
};
\addlegendentry{clb}

\end{axis}
\end{tikzpicture}%
\caption{\figcap{Correlation between activation-derived F0 and reference F0. Plateaus are observed with increasing correlation.}}
\label{fig:f0}
\vspace{-10pt}
\end{figure}
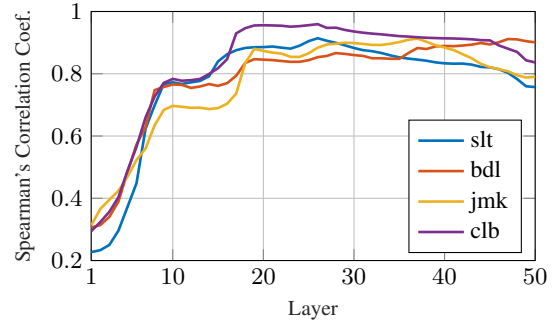
\begin{figure}
\centering
%
%
\begin{tikzpicture}

\begin{axis}[%
width=1.0in,
height=1.2in,
at={(0in,0.2in)},
scale only axis,
point meta min=0.5,
point meta max=1,
axis on top,
xmin=1,
xmax=50,
xlabel near ticks,
xtick = {1, 10, 20, 30, 40, 50},
xlabel style={font=\color{white!15!black}},
xlabel={\footnotesize{Layer}},
ymin=0,
ymax=8,
ylabel near ticks,
ylabel style={font=\color{white!15!black}},
ylabel={\footnotesize{Channel Frequency (kHz)}},
axis background/.style={fill=white},
title style={font=\bfseries},
title={slt},
xmajorgrids,
grid style = {dashed},
]
\addplot [forget plot] graphics [xmin=1, xmax=50, ymin=0, ymax=8] {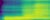};
\end{axis}

\begin{axis}[%
width=1.0in,
height=1.2in,
at={(1.2in,0.2in)},
scale only axis,
point meta min=0.5,
point meta max=1,
axis on top,
xmin=1,
xmax=50,
xlabel near ticks,
xtick = {1, 10, 20, 30, 40, 50},
xlabel style={font=\color{white!15!black}},
xlabel={\footnotesize{Layer}},
ymin=0,
ymax=8,
ylabel near ticks,
axis background/.style={fill=white},
title style={font=\bfseries},
title={bdl},
xmajorgrids,
grid style = {dashed},
colormap={mymap}{[1pt] rgb(0pt)=(0.267004,0.00487433,0.329415); rgb(1pt)=(0.272652,0.0258457,0.353367); rgb(2pt)=(0.277106,0.0509139,0.376236); rgb(3pt)=(0.280356,0.0742015,0.397901); rgb(4pt)=(0.28239,0.0959536,0.418251); rgb(5pt)=(0.283205,0.116893,0.437179); rgb(6pt)=(0.282809,0.13735,0.454596); rgb(7pt)=(0.281231,0.15748,0.470434); rgb(8pt)=(0.278516,0.177348,0.484654); rgb(9pt)=(0.274736,0.196969,0.49725); rgb(10pt)=(0.269982,0.21633,0.508255); rgb(11pt)=(0.264369,0.235405,0.517732); rgb(12pt)=(0.258026,0.254162,0.52578); rgb(13pt)=(0.251099,0.272573,0.532522); rgb(14pt)=(0.243733,0.29062,0.538097); rgb(15pt)=(0.236073,0.308291,0.542652); rgb(16pt)=(0.228263,0.325586,0.546335); rgb(17pt)=(0.220425,0.342517,0.549287); rgb(18pt)=(0.212667,0.359102,0.551635); rgb(19pt)=(0.205079,0.375366,0.553493); rgb(20pt)=(0.197722,0.391341,0.554953); rgb(21pt)=(0.190631,0.407061,0.556089); rgb(22pt)=(0.183819,0.422564,0.556952); rgb(23pt)=(0.177272,0.437886,0.557576); rgb(24pt)=(0.170958,0.453063,0.557974); rgb(25pt)=(0.164833,0.46813,0.558143); rgb(26pt)=(0.158845,0.483117,0.558059); rgb(27pt)=(0.152951,0.498053,0.557685); rgb(28pt)=(0.147132,0.512959,0.556973); rgb(29pt)=(0.141402,0.527854,0.555864); rgb(30pt)=(0.135833,0.54275,0.554289); rgb(31pt)=(0.130582,0.557652,0.552176); rgb(32pt)=(0.125898,0.572563,0.549445); rgb(33pt)=(0.122163,0.587476,0.546023); rgb(34pt)=(0.119872,0.602382,0.541831); rgb(35pt)=(0.119627,0.617266,0.536796); rgb(36pt)=(0.122046,0.632107,0.530848); rgb(37pt)=(0.127668,0.646882,0.523924); rgb(38pt)=(0.136835,0.661563,0.515967); rgb(39pt)=(0.149643,0.67612,0.506924); rgb(40pt)=(0.165967,0.690519,0.496752); rgb(41pt)=(0.185538,0.704725,0.485412); rgb(42pt)=(0.20803,0.718701,0.472873); rgb(43pt)=(0.233127,0.732406,0.459106); rgb(44pt)=(0.260531,0.745802,0.444096); rgb(45pt)=(0.290001,0.758846,0.427826); rgb(46pt)=(0.32133,0.771498,0.410293); rgb(47pt)=(0.354355,0.783714,0.391488); rgb(48pt)=(0.38893,0.795453,0.371421); rgb(49pt)=(0.424933,0.806674,0.350099); rgb(50pt)=(0.462247,0.817338,0.327545); rgb(51pt)=(0.500754,0.827409,0.303799); rgb(52pt)=(0.540337,0.836858,0.278917); rgb(53pt)=(0.580861,0.845663,0.253001); rgb(54pt)=(0.622171,0.853816,0.226224); rgb(55pt)=(0.664087,0.861321,0.198879); rgb(56pt)=(0.706404,0.868206,0.171495); rgb(57pt)=(0.748885,0.874522,0.145038); rgb(58pt)=(0.791273,0.880346,0.121291); rgb(59pt)=(0.833302,0.88578,0.103326); rgb(60pt)=(0.874718,0.890945,0.0953508); rgb(61pt)=(0.915296,0.895973,0.10047); rgb(62pt)=(0.95484,0.901006,0.117876); rgb(63pt)=(0.993248,0.906157,0.143936)},
colorbar,
colorbar style={
  width = 0.15in,
  title = Corr.,
  ytick = {0.5, 0.6, 0.7, 0.8, 0.9, 1.0},
  ticklabel style = {font=\footnotesize}
}
]
\addplot [forget plot] graphics [xmin=1, xmax=50, ymin=0, ymax=8] {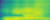};
\end{axis}
\end{tikzpicture}%
\caption{\figcap{Correlation between activation-derived band energy and reference energy for speaker \textit{slt} (female) and \textit{bdl} (male).}}
\label{fig:band-energy}
\vspace{-10pt}
\end{figure}

\subsection{Band Energy Regression from Hidden Layers}

Similar to the experiment on log F0, the quality of band energy (in dB) reconstructed from the activations is evaluated in a similar way but repeated for different frequency bands. With the exception of the first few layers below the receptive field limit, high correlation (greater than $0.9$) is consistently observed at all layers for frequencies below 1 kHz. The pattern above 1 kHz is more complicated with a low initial correlation and multiple plateaus spanning across convolution block boundaries and rises are often observed in the middle of a block (e.g. layer 15, 25, and 35.) At this point it is unclear why the contents in mid-high frequency, covering most of the speech formants and spectral cues important to consonant perception \cite{li-2010} are not identified until a much later stage in the network.

\subsection{Magnitude Spectra Regression from Hidden Layers}

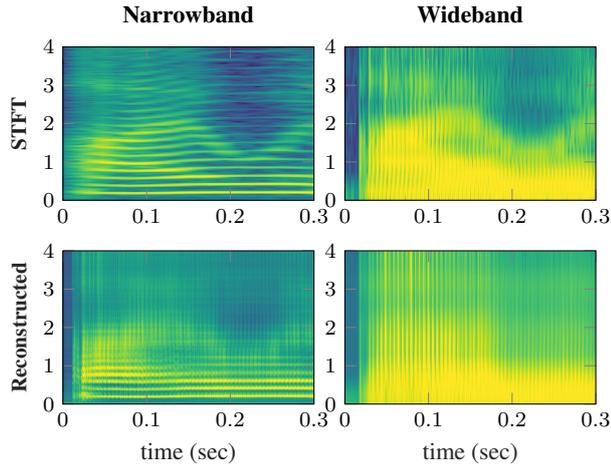
\begin{figure}
\centering
%
%
\definecolor{mycolor1}{rgb}{0.00000,0.44700,0.74100}%
\begin{tikzpicture}

\begin{axis}[%
width=1.3in,
height=0.8in,
at={(0.84in,1.5in)},
scale only axis,
point meta min=-7,
point meta max=4,
unbounded coords=jump,
axis on top,
xmin=0,
xmax=0.3,
ymin=0,
ymax=4,
ticklabel style = {font=\footnotesize},
ylabel near ticks,
ylabel style={font=\color{white!12!black}},
ylabel={\footnotesize \textbf{STFT}},
axis background/.style={fill=white},
title style={font=\bfseries},
title={Narrowband}
]
\addplot [color=mycolor1, mark=*, mark options={solid, mycolor1}, forget plot]
  table[row sep=crcr]{%
1	nan\\
};
\addplot [forget plot] graphics [xmin=-0.00025, xmax=0.29975, ymin=-0.0078125, ymax=7.9921875] {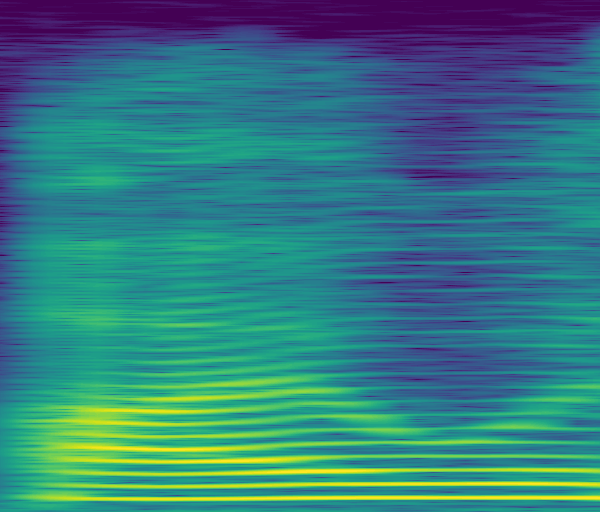};
\end{axis}

\begin{axis}[%
width=1.3in,
height=0.8in,
at={(0.84in,0.436in)},
scale only axis,
point meta min=-7,
point meta max=4,
unbounded coords=jump,
axis on top,
xmin=0,
xmax=0.3,
ticklabel style = {font=\footnotesize},
xlabel near ticks,
xlabel style={font=\color{white!12!black}},
xlabel={time (sec)},
ymin=0,
ymax=4,
ylabel near ticks,
ylabel style={font=\color{white!15!black}},
ylabel={\footnotesize \textbf{Reconstructed}},
axis background/.style={fill=white}
]
\addplot [color=mycolor1, mark=*, mark options={solid, mycolor1}, forget plot]
  table[row sep=crcr]{%
1	nan\\
};
\addplot [forget plot] graphics [xmin=-0.00025, xmax=0.29975, ymin=-0.0078125, ymax=7.9921875] {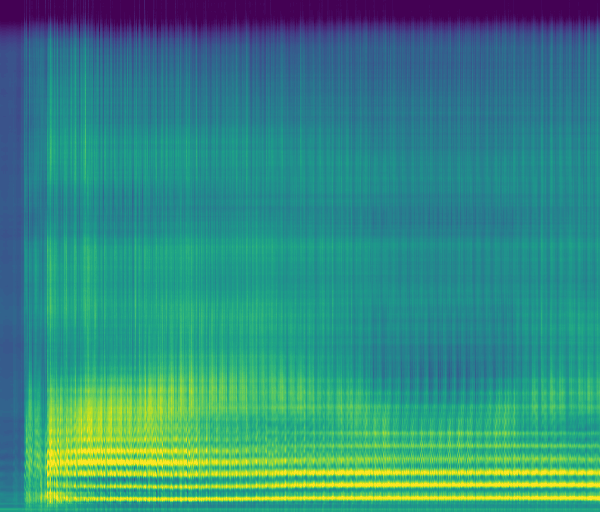};
\end{axis}

\begin{axis}[%
width=1.3in,
height=0.8in,
at={(2.3in,1.5in)},
scale only axis,
point meta min=-10,
point meta max=1,
unbounded coords=jump,
axis on top,
xmin=0,
xmax=0.3,
ymin=0,
ymax=4,
axis background/.style={fill=white},
ticklabel style = {font=\footnotesize},
title style={font=\bfseries},
title={Wideband}
]
\addplot [color=mycolor1, mark=*, mark options={solid, mycolor1}, forget plot]
  table[row sep=crcr]{%
1	nan\\
};
\addplot [forget plot] graphics [xmin=-0.00025, xmax=0.29975, ymin=-0.128780241935484, ymax=8.11315524193548] {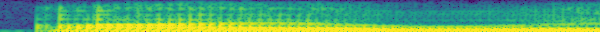};
\end{axis}

\begin{axis}[%
width=1.3in,
height=0.8in,
at={(2.3in,0.436in)},
scale only axis,
point meta min=-10,
point meta max=1,
unbounded coords=jump,
axis on top,
xmin=0,
xmax=0.3,
ticklabel style = {font=\footnotesize},
xlabel near ticks,
xlabel style={font=\color{white!15!black}},
xlabel={time (sec)},
ymin=0,
ymax=4,
axis background/.style={fill=white}
]
\addplot [color=mycolor1, mark=*, mark options={solid, mycolor1}, forget plot]
  table[row sep=crcr]{%
1	nan\\
};
\addplot [forget plot] graphics [xmin=-0.00025, xmax=0.29975, ymin=-0.128780241935484, ymax=8.11315524193548] {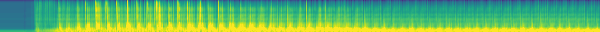};
\end{axis}
\end{tikzpicture}%
\caption{\figcap{Examples of spectrograms and their activation-derived versions on speaker \textit{slt} (female), using outputs from layer 25. Spectral-temporal details are selectively preserved.}}
\label{fig:spec}
\vspace{-10pt}
\end{figure}

Frequency-layer plots on the task of STFT magnitude spectrum reconstruction closely resemble the results shown in Figure~\ref{fig:band-energy} and are thus not included in this paper. Examples of STFT spectrograms and their approximated versions are given in Figure~\ref{fig:spec}. Although the spectrum tends to be over-smoothed at higher frequencies, the reconstruction derived from a linear transform of activations preserves the first 5 or 6 harmonics in the narrowband case and some excitation structures can be seen in the wideband case. Note that both spectrograms are derived from the same set of 64 units in layer 25.

Manual inspection shows that the regression weights mapping activation vectors to log magnitude bins, after proper reordering, visually resemble the harmonic basis for discrete cosine transform. It is thus intriguing to think of the activations as a cepstral representation of speech. However, by taking the Fourier transform of regression weights, we find that the basis is not harmonic. There is yet no evidence supporting the existence of an assignment of interpretable meaning to each of the units.

\subsection{Discussions and Further Analysis}

Experiments in the previous sections have shown that the activations in each dilated convolution layer encode information about both baseband (i.e. slowly varying) and wideband (i.e. rapidly varying) features. The ability to separate the two components via a linear combination learned by regression implies a certain degree of degeneracy in the matrix of activations.

\begin{figure}
\centering
%
%
\begin{tikzpicture}

\begin{axis}[%
width=1in,
height=1.2in,
at={(0in,0.2in)},
scale only axis,
point meta min=2.5,
point meta max=6.5,
axis on top,
xmin=1,
xmax=64,
xtick = {1, 16, 32, 48, 64},
xlabel near ticks,
xlabel style={font=\color{white!15!black}},
xlabel={\footnotesize{Sorted Units}},
y dir=reverse,
ymin=1,
ymax=50,
ytick = {1, 10, 20, 30, 40, 50},
ylabel near ticks,
ylabel style={font=\color{white!15!black}},
ylabel={\footnotesize{Layer}},
axis background/.style={fill=white},
title style={font=\bfseries},
title={Baseband},
ymajorgrids,
grid style = {dashed}
]
\addplot [forget plot] graphics [xmin=0.5, xmax=64.5, ymin=0.5, ymax=50.5] {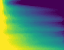};
\end{axis}

\begin{axis}[%
width=1in,
height=1.2in,
at={(1.3in,0.2in)},
scale only axis,
point meta min=2.5,
point meta max=6.5,
axis on top,
xmin=1,
xmax=64,
xtick = {1, 16, 32, 48, 64},
xlabel near ticks,
xlabel style={font=\color{white!15!black}},
xlabel={\footnotesize{Sorted Units}},
y dir=reverse,
ymin=1,
ymax=50,
ytick = {1, 10, 20, 30, 40, 50},
axis background/.style={fill=white},
title style={font=\bfseries},
title={Wideband},
ymajorgrids,
grid style = {dashed},
colormap={mymap}{[1pt] rgb(0pt)=(0.267004,0.00487433,0.329415); rgb(1pt)=(0.272652,0.0258457,0.353367); rgb(2pt)=(0.277106,0.0509139,0.376236); rgb(3pt)=(0.280356,0.0742015,0.397901); rgb(4pt)=(0.28239,0.0959536,0.418251); rgb(5pt)=(0.283205,0.116893,0.437179); rgb(6pt)=(0.282809,0.13735,0.454596); rgb(7pt)=(0.281231,0.15748,0.470434); rgb(8pt)=(0.278516,0.177348,0.484654); rgb(9pt)=(0.274736,0.196969,0.49725); rgb(10pt)=(0.269982,0.21633,0.508255); rgb(11pt)=(0.264369,0.235405,0.517732); rgb(12pt)=(0.258026,0.254162,0.52578); rgb(13pt)=(0.251099,0.272573,0.532522); rgb(14pt)=(0.243733,0.29062,0.538097); rgb(15pt)=(0.236073,0.308291,0.542652); rgb(16pt)=(0.228263,0.325586,0.546335); rgb(17pt)=(0.220425,0.342517,0.549287); rgb(18pt)=(0.212667,0.359102,0.551635); rgb(19pt)=(0.205079,0.375366,0.553493); rgb(20pt)=(0.197722,0.391341,0.554953); rgb(21pt)=(0.190631,0.407061,0.556089); rgb(22pt)=(0.183819,0.422564,0.556952); rgb(23pt)=(0.177272,0.437886,0.557576); rgb(24pt)=(0.170958,0.453063,0.557974); rgb(25pt)=(0.164833,0.46813,0.558143); rgb(26pt)=(0.158845,0.483117,0.558059); rgb(27pt)=(0.152951,0.498053,0.557685); rgb(28pt)=(0.147132,0.512959,0.556973); rgb(29pt)=(0.141402,0.527854,0.555864); rgb(30pt)=(0.135833,0.54275,0.554289); rgb(31pt)=(0.130582,0.557652,0.552176); rgb(32pt)=(0.125898,0.572563,0.549445); rgb(33pt)=(0.122163,0.587476,0.546023); rgb(34pt)=(0.119872,0.602382,0.541831); rgb(35pt)=(0.119627,0.617266,0.536796); rgb(36pt)=(0.122046,0.632107,0.530848); rgb(37pt)=(0.127668,0.646882,0.523924); rgb(38pt)=(0.136835,0.661563,0.515967); rgb(39pt)=(0.149643,0.67612,0.506924); rgb(40pt)=(0.165967,0.690519,0.496752); rgb(41pt)=(0.185538,0.704725,0.485412); rgb(42pt)=(0.20803,0.718701,0.472873); rgb(43pt)=(0.233127,0.732406,0.459106); rgb(44pt)=(0.260531,0.745802,0.444096); rgb(45pt)=(0.290001,0.758846,0.427826); rgb(46pt)=(0.32133,0.771498,0.410293); rgb(47pt)=(0.354355,0.783714,0.391488); rgb(48pt)=(0.38893,0.795453,0.371421); rgb(49pt)=(0.424933,0.806674,0.350099); rgb(50pt)=(0.462247,0.817338,0.327545); rgb(51pt)=(0.500754,0.827409,0.303799); rgb(52pt)=(0.540337,0.836858,0.278917); rgb(53pt)=(0.580861,0.845663,0.253001); rgb(54pt)=(0.622171,0.853816,0.226224); rgb(55pt)=(0.664087,0.861321,0.198879); rgb(56pt)=(0.706404,0.868206,0.171495); rgb(57pt)=(0.748885,0.874522,0.145038); rgb(58pt)=(0.791273,0.880346,0.121291); rgb(59pt)=(0.833302,0.88578,0.103326); rgb(60pt)=(0.874718,0.890945,0.0953508); rgb(61pt)=(0.915296,0.895973,0.10047); rgb(62pt)=(0.95484,0.901006,0.117876); rgb(63pt)=(0.993248,0.906157,0.143936)},
colorbar,
colorbar style = {width = 0.13in, at = {(2.3in, 1.4in)}}
]
\addplot [forget plot] graphics [xmin=0.5, xmax=64.5, ymin=0.5, ymax=50.5] {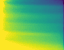};
\end{axis}
\end{tikzpicture}%
\caption{\figcap{Sorted log singular values of activation matrices on speaker \textit{slt} (female) reveal a periodic structure across convolution blocks. The pattern is replicable on other speakers as well.}}
\label{fig:svd}
\vspace{-10pt}
\end{figure}

Previously singular value decomposition (SVD) has been employed to explore the redundancy in learned weights \cite{xue-2013}; in this study, we apply SVD on low-pass filtered and high-pass filtered activations, and visualize the sorted singular values to characterize the resource allocation in the activation space. The activations are processed by a 2nd-order Linkwitz-Riley crossover filter \cite{linkwitz-1976} with an $80$ Hz cutoff frequency and examples for baseband/wideband singular values taken from one of the speakers are shown in Figure~\ref{fig:svd}.

It is immediately observable that a complementary and periodic (along layers) pattern exists across baseband and wideband singular values. The tails of baseband singular values are typically longer near the boundary of a convolution block (layer 10, 20, 30, 40 and 50) while the tails of wideband singular values are longer, though less apparently, near the middle of a convolution block (layer 5, 15, 25, and 35).

\begin{figure}
\centering
%
%
\begin{tikzpicture}

\begin{axis}[%
width=1in,
height=1.2in,
at={(0in,0.2in)},
scale only axis,
point meta min=0,
point meta max=1,
axis on top,
xmin=1,
xmax=128,
xtick = {1, 32, 64, 96, 128},
xlabel near ticks,
xlabel style={font=\color{white!15!black}},
xlabel={\footnotesize{Sorted Filter/Gate Indices}},
y dir=reverse,
ymin=1,
ymax=50,
ytick = {1, 10, 20, 30, 40, 50},
ylabel near ticks,
ylabel style={font=\color{white!15!black}},
ylabel={\footnotesize{Layer}},
axis background/.style={fill=white},
title style={font=\bfseries},
title={slt},
ymajorgrids,
grid style = {dashed}
]
\addplot [forget plot] graphics [xmin=0.5, xmax=128.5, ymin=0.5, ymax=49.5] {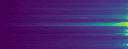};
\end{axis}

\begin{axis}[%
width=1in,
height=1.2in,
at={(1.3in,0.2in)},
scale only axis,
point meta min=0,
point meta max=1,
axis on top,
xmin=1,
xmax=128,
xtick = {1, 32, 64, 96, 128},
xlabel near ticks,
xlabel style={font=\color{white!15!black}},
xlabel={\footnotesize{Sorted Filter/Gate Indices}},
y dir=reverse,
ymin=1,
ymax=50,
ytick = {1, 10, 20, 30, 40, 50},
axis background/.style={fill=white},
title style={font=\bfseries},
title={bdl},
ymajorgrids,
grid style = {dashed},
colormap={mymap}{[1pt] rgb(0pt)=(0.267004,0.00487433,0.329415); rgb(1pt)=(0.272652,0.0258457,0.353367); rgb(2pt)=(0.277106,0.0509139,0.376236); rgb(3pt)=(0.280356,0.0742015,0.397901); rgb(4pt)=(0.28239,0.0959536,0.418251); rgb(5pt)=(0.283205,0.116893,0.437179); rgb(6pt)=(0.282809,0.13735,0.454596); rgb(7pt)=(0.281231,0.15748,0.470434); rgb(8pt)=(0.278516,0.177348,0.484654); rgb(9pt)=(0.274736,0.196969,0.49725); rgb(10pt)=(0.269982,0.21633,0.508255); rgb(11pt)=(0.264369,0.235405,0.517732); rgb(12pt)=(0.258026,0.254162,0.52578); rgb(13pt)=(0.251099,0.272573,0.532522); rgb(14pt)=(0.243733,0.29062,0.538097); rgb(15pt)=(0.236073,0.308291,0.542652); rgb(16pt)=(0.228263,0.325586,0.546335); rgb(17pt)=(0.220425,0.342517,0.549287); rgb(18pt)=(0.212667,0.359102,0.551635); rgb(19pt)=(0.205079,0.375366,0.553493); rgb(20pt)=(0.197722,0.391341,0.554953); rgb(21pt)=(0.190631,0.407061,0.556089); rgb(22pt)=(0.183819,0.422564,0.556952); rgb(23pt)=(0.177272,0.437886,0.557576); rgb(24pt)=(0.170958,0.453063,0.557974); rgb(25pt)=(0.164833,0.46813,0.558143); rgb(26pt)=(0.158845,0.483117,0.558059); rgb(27pt)=(0.152951,0.498053,0.557685); rgb(28pt)=(0.147132,0.512959,0.556973); rgb(29pt)=(0.141402,0.527854,0.555864); rgb(30pt)=(0.135833,0.54275,0.554289); rgb(31pt)=(0.130582,0.557652,0.552176); rgb(32pt)=(0.125898,0.572563,0.549445); rgb(33pt)=(0.122163,0.587476,0.546023); rgb(34pt)=(0.119872,0.602382,0.541831); rgb(35pt)=(0.119627,0.617266,0.536796); rgb(36pt)=(0.122046,0.632107,0.530848); rgb(37pt)=(0.127668,0.646882,0.523924); rgb(38pt)=(0.136835,0.661563,0.515967); rgb(39pt)=(0.149643,0.67612,0.506924); rgb(40pt)=(0.165967,0.690519,0.496752); rgb(41pt)=(0.185538,0.704725,0.485412); rgb(42pt)=(0.20803,0.718701,0.472873); rgb(43pt)=(0.233127,0.732406,0.459106); rgb(44pt)=(0.260531,0.745802,0.444096); rgb(45pt)=(0.290001,0.758846,0.427826); rgb(46pt)=(0.32133,0.771498,0.410293); rgb(47pt)=(0.354355,0.783714,0.391488); rgb(48pt)=(0.38893,0.795453,0.371421); rgb(49pt)=(0.424933,0.806674,0.350099); rgb(50pt)=(0.462247,0.817338,0.327545); rgb(51pt)=(0.500754,0.827409,0.303799); rgb(52pt)=(0.540337,0.836858,0.278917); rgb(53pt)=(0.580861,0.845663,0.253001); rgb(54pt)=(0.622171,0.853816,0.226224); rgb(55pt)=(0.664087,0.861321,0.198879); rgb(56pt)=(0.706404,0.868206,0.171495); rgb(57pt)=(0.748885,0.874522,0.145038); rgb(58pt)=(0.791273,0.880346,0.121291); rgb(59pt)=(0.833302,0.88578,0.103326); rgb(60pt)=(0.874718,0.890945,0.0953508); rgb(61pt)=(0.915296,0.895973,0.10047); rgb(62pt)=(0.95484,0.901006,0.117876); rgb(63pt)=(0.993248,0.906157,0.143936)},
colorbar,
colorbar style = {width = 0.13in, at = {(2.3in, 1.4in)}}
]
\addplot [forget plot] graphics [xmin=0.5, xmax=128.5, ymin=0.5, ymax=49.5] {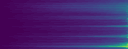};
\end{axis}
\end{tikzpicture}%
\caption{\figcap{Explicit correlations between filter/gating results and F0 on speaker \textit{slt} (female) and \textit{bdl} (male).}}
\label{fig:f0corr}
\vspace{-10pt}
\end{figure}

The said phenomenon encourages a theory, furthering the claim made in section~\ref{sec:logf0}, that WaveNet performs analysis on wideband activation components (e.g. reconstructed waveform) in the middle portion of each convolution block and stores the results into the baseband at the end of the block. To verify this theory, we look for evidences that WaveNet \textit{explicitly} computes F0 from the activations in the middle layers. This is done by evaluating the terms inside the non-linear functions in (\ref{eq:wavenet-layer}) and computing the Spearman's correlation coefficient with log F0 extracted from the input speech. Interestingly, the results in Figure~\ref{fig:f0corr} are in well-agreement with the theory and a maximal Spearman's correlation coefficient above 0.8 is observed for both speakers, despite that the WaveNet model is trained in an unsupervised setup.

\section{Conclusions}

We have evaluated the capability of unsupervisedly trained WaveNets to encode speech features in the activations. The findings point to a compact representation relating to spectral-temporal details of the speech signal, formed by means of iteratively analyzing the rapidly changing signals and storing the results in a temporally smooth component. This iterative behavior is encouraged by the stacking of convolution blocks with increasing dilation factors.

Although introducing architectural changes with the goal of quality improvement is beyond the scope of this study, the findings suggest several possibilities worth exploring in future studies. It is tempting to augment the dilated convolution with recurrent low-pass filters to better separate baseband features from the noisy activations, reducing the (frequency-wise) degeneracy of activation matrices. In addition, we expect an increase in the frequency resolution of reconstructed spectrograms (Figure~\ref{fig:spec}) if the layer size can be larger.

\section{Acknowledgments}

The author would like to thank Vassilis Tsiaras from the University of Crete for providing a WaveNet implementation \cite{adiga-2018} and to thank Shotaro Ikeda from the University of Illinois for his supports on Tensorflow programming.

\bibliographystyle{IEEEtran}

\bibliography{mybib}

\end{document}